\documentclass[a4paper,11pt]{article} 
\usepackage[margin=3cm]{geometry}
\usepackage{amstext,amsmath,amssymb}
\usepackage{blindtext}
\usepackage{enumitem}
\usepackage{changepage, environ}
\PassOptionsToPackage{pass,showframe}{ge	ometry}
\usepackage{float}
\usepackage[utf8]{inputenc}
\usepackage[T1]{fontenc}
\usepackage[autostyle]{csquotes}
\usepackage{dirtytalk}
\usepackage{graphicx} 
\usepackage{caption}
\usepackage[english]{babel}
\usepackage{xcolor}
\usepackage{breakurl}
\usepackage[breaklinks]{hyperref}
\usepackage{setspace}
\hypersetup{colorlinks=true, pdfstartview=FitV, linkcolor=blue, citecolor=black, plainpages=false, pdfpagelabels=true, urlcolor=blue}
\usepackage{xurl}
\usepackage{scalerel}
\usepackage{tikz}
\usepackage{multicol}
\usepackage{blindtext}
\usepackage{abstract}
\usepackage{sectsty}
\usepackage{secdot}
\usepackage{fancyhdr}
\usepackage{secdot}
\usepackage{blindtext}
\usepackage{academicons}
\usepackage{graphicx}
\graphicspath{ {./images/} }

\PassOptionsToPackage{dvipsnames}{xcolor}
\usetikzlibrary{decorations.text}
\usepackage{tikz}
\usepackage[nottoc,numbib]{tocbibind}
\usepackage{setspace}

\IfFileExists{upquote.sty}{\usepackage{upquote}}{}
\IfFileExists{microtype.sty}{
  \usepackage[]{microtype}
  \UseMicrotypeSet[protrusion]{basicmath} 
}{}
\makeatletter
\@ifundefined{KOMAClassName}{
  \IfFileExists{parskip.sty}{%
    \usepackage{parskip}
  }{
    \setlength{\parindent}{0pt}
    \setlength{\parskip}{6pt plus 2pt minus 1pt}}
}{
  \KOMAoptions{parskip=half}}
\makeatother
\usepackage{longtable,booktabs,array}
\usepackage{calc} 

\title{\textbf{AI Act and Large Language Models (LLMs):\\ When critical issues and privacy impact require human and ethical oversight.}}

\usepackage[multiple]{footmisc}
\usepackage{bigfoot}

\DeclareNewFootnote{AAffil}[arabic]
\DeclareNewFootnote{ANote}[fnsymbol]

\NewEnviron{myquote}[1][\relax]{
  \begin{center}
  \begin{adjustwidth}{50pt}{50pt}
  \setbox0=\hbox{\itshape\BODY}%
  \begin{itshape}
  \ifdim\wd0>\linewidth\relax\BODY\else\hfil\BODY\fi
  \end{itshape}
  \end{adjustwidth}
  \end{center}  
  \ifx\relax#1\relax\else
    \vspace{-0.15cm}
    \begin{adjustwidth}{0pt}{20pt}
    \flushright { \small -\hspace{2pt}#1\hspace{2pt}- }
    \end{adjustwidth}
    \vspace{0.3cm}
  \fi
}

\author{
  Nicola Fabiano\thanks{\href{https://www.fabiano.law/en/page/about/}{Studio Legale Fabiano (Italy)} - Affiliation: \textit{International Institute of Informatics and Systemics (IIIS) - USA}}
}

\date{March 30, 2024}

\begin{document}
\maketitle

\begin{abstract}

The imposing evolution of artificial intelligence systems and, specifically, of Large Language Models (LLM) makes it necessary to carry out assessments of their level of risk and the impact they may have in the area of privacy, personal data protection and at an ethical level, especially on the weakest and most vulnerable. This contribution addresses human oversight, ethical oversight, and privacy impact assessment. 

\end{abstract}


\begin{spacing}{0.2}
\tableofcontents
\end{spacing}
\newpage

\section{Introduction}

On March 13, 2024, the European Parliament approved the final version of
the European Artificial Intelligence Act (\textbf{AI Act}), and its
publication in the Official Journal of the European Union is awaited.
The AI Act is a long text comprising 180 recitals, XIII chapters with
113 articles, and XIII annexes.

It is an essential legal framework for AI and the first comprehensive
legislation on AI.

\section{The implementation of the AI Act}

The European Artificial Intelligence Act, according to Article 113
(Entry into force and enforcement), will enter into \textbf{force 20
days} after its publication in the Official Journal of the European
Union and will then become fully \textbf{applicable} in the following
four main stages:

\begin{longtable}[]{@{}
  >{\raggedright\arraybackslash}p{(\columnwidth - 4\tabcolsep) * \real{0.1510}}
  >{\raggedright\arraybackslash}p{(\columnwidth - 4\tabcolsep) * \real{0.1887}}
  >{\raggedright\arraybackslash}p{(\columnwidth - 4\tabcolsep) * \real{0.6604}}@{}}
\toprule()
\begin{minipage}[b]{\linewidth}\raggedright
\textbf{Phases}
\end{minipage} & \begin{minipage}[b]{\linewidth}\raggedright
\textbf{When}
\end{minipage} & \begin{minipage}[b]{\linewidth}\raggedright
\textbf{Rules of the DI Act}
\end{minipage} \\
\midrule()
\endhead
\textbf{1} & \textbf{6 months} &
\begin{minipage}[t]{\linewidth}\raggedright
\begin{itemize}
\item
  Chapter I (General Provisions)
\item
  Chapter II (Prohibited Artificial Intelligence Practices)
\end{itemize}
\end{minipage} \\
\textbf{2} & \textbf{12 months} &
\begin{minipage}[t]{\linewidth}\raggedright
\begin{itemize}
\item
  Chapter III - Section 4 (Notifying authorities and notified bodies)
\item
  Chapter V (AI Models for General Purposes)
\item
  Chapter VII (Governance)
\item
  Chapter XII (Penalties), except for Article 101 (Financial penalties
  for providers of general purpose AI models)
\end{itemize}
\end{minipage} \\
\textbf{3} & \textbf{24 months} & \textbf{All regulations of the DI Act
become applicable} \\
\textbf{4} & \textbf{36 months} &
\begin{minipage}[t]{\linewidth}\raggedright
\begin{itemize}
\item
  Article 6(1) (Classification rules for high-risk AI systems) and
  corresponding obligations
\end{itemize}
\end{minipage} \\
\bottomrule()
\end{longtable}

\section{The definition of artificial intelligence systems}

Preliminarily, given its relevance in the recently approved regulatory
framework, it is worth briefly commenting on the definition of AI
systems.

As outlined at international conferences \cite{AMALEA23}-\cite{APPIS24}, three different versions were presented during the legislative process: one by the European Commission, one by the Council of Europe, and one by the European Parliament.

The three proposals discussed during the trilogue, which kept many
people glued to their devices to follow the news from December 7, 2023,
ended on December 9, 2023, after a grueling three-day marathon
discussion.

On February 13, 2024, the European Parliament\textquotesingle s LIBE and
IMCO committees adopted the final text of the EU Artificial Intelligence
Act in a joint vote by an overwhelming majority (71 votes in favor, 8
votes against, and 7 abstentions). Council negotiators also approved the
text in February 2024.

The three versions of the definition of artificial intelligence system
that have been discussed are as follows:

\begin{myquote}
\emph{\textbf{Article 3 (EU Commission Proposal - April 21, 2021)}}
\end{myquote}

\begin{quote}

\emph{\textquotesingle{}\textbf{artificial intelligence
system}\textquotesingle{} (AI system) means software that is developed
with one or more of the techniques and approaches listed in Annex I and
can, for a given set of human-defined objectives, generate outputs such
as content, predictions, recommendations, or decisions influencing the
environments they interact with;}
\end{quote}
\hfill
\break

\begin{myquote}
\emph{\textbf{Article 3 (Council of the European Union - November 25, 2022)}}
\end{myquote}

\begin{quote}
	
\emph{\textquotesingle{}\textbf{artificial intelligence
system}\textquotesingle{} (AI system) means a system that is designed to
operate with elements of autonomy and that, based on machine and/or
human-provided data and inputs, infers how to achieve a given set of
objectives using machine learning and/or logic- and knowledge based
approaches, and produces system-generated outputs such as content
(generative AI systems), predictions, recommendations or decisions,
influencing the environments with which the AI system interacts;}
\end{quote}
\hfill
\break

\begin{myquote}
\emph{\textbf{Article 3 (EU Parliament - June 14, 2023)}}
\end{myquote}

\begin{quote}

\emph{\textquotesingle{}\textbf{artificial intelligence
system}\textquotesingle{} (AI system) means a machine-based system that
is designed to operate with varying levels of autonomy and that can, for
explicit or implicit objectives, generate outputs such as predictions,
recommendations, or decisions, that influence physical or virtual
environments.}
\end{quote}
\hfill
\break

It seemed that the EU legislature would adopt the definition provided by
the OECD contained in the latest version of the Council Recommendation
on Artificial Intelligence \cite{OECD}, amended in November 2023\footnote{The OECD \href{https://legalinstruments.oecd.org/en/instruments/oecd-legal-0449}{Recommendation of the Council on Artificial
  Intelligence} provides the following definition of Artificial Intelligence System: ``\textit{AI system: An AI system is a machine-based system that, for explicit or implicit objectives, infers, from the input it receives, how to generate outputs such as predictions, content, recommendations, or decisions that can influence physical or virtual environments. Different AI systems vary in their levels of autonomy and adaptiveness after deployment."}.}.

However, the European legislature has adopted a definition of an AI
system which is very different and broader than that provided by the
OECD and is as follows:

\begin{quote}
\emph{\textquotesingle{}\textbf{AI system}\textquotesingle{} means a
machine-based system designed to operate with varying levels of
autonomy, that may exhibit adaptiveness after deployment and that, for
explicit or implicit objectives, infers, from the input it receives, how
to generate outputs such as predictions, content, recommendations, or
decisions that can influence physical or virtual environments.}
\end{quote}

The above definition makes some clarifications appropriate. In
particular, the term "artificial intelligence" seems to be attributed to
John McCarthy of the MIT (Massachusetts Institute of Technology) \cite{coe},
who used it during an academic conference held at Dartmouth College from
June 19 to August 16, 1956. However, there has been much debate over the
expression "artificial intelligence" in the years to follow, trying to
identify a definition. In particular, what was stated by Stuart Russel
and Peter Norvig\footnote{The authors state, ``\emph{Historically, researchers have pursued several different versions of AI. Some have defined intelligence in terms fidelity to human performance, while others prefer an abstract, formal definition of intelligence called rationality–loosely speaking, doing the \textquotesingle ‘right thing’ \textquotesingle{}. 
The subject matter itself also varies: some consider intelligence to be a property of internal thought processes and reasoning, while others focus on intelligent behavior, an external characterization. 
From these two dimensions – human vs. rational and thought vs behavior – there are four possible combinations, and there have been adherents and research programs for all four."}.}  in their book entitled ``\emph{Artiﬁcial Intelligence. A
Modern Approach - Fourth Edition}" (2020) \cite{russel} shows that artificial intelligence cannot be defined.

Therefore, it is correct to identify the definition of ``artificial intelligence system", given that the expression ``artificial intelligence" refers to a technology that represents the whole of which Machine Learning (ML), Deep Learning (DL), Natural Language Processing (NLP) and Large Language Models (LLM), Computer Vision, Robotics, Neural Networks, and Cognitive computing are part as of subsets.

\section{AI Act and high-risk systems}

The European Artificial Intelligence Act (AI Act) regulates high-risk
systems in Chapter III.

For this paper, we will focus on the ``\emph{classification of AI systems
as \textquotesingle high-risk\textquotesingle{}} and the ``\emph{requirements for high-risk AI systems"}, which are covered in Section 1 and Section 2, respectively.

The starting point is Section 1, which governs the ``\emph{classification
rules for high-risk AI systems}", specifying that to classify an AI system as high-risk-regardless of whether it has been placed on the market or put into service \textbf{both of the} conditions outlined in Article 6(1)(a) and (b) must be met, namely:

\begin{enumerate}
\def\labelenumi{(\alph{enumi})}
\item
\textit{the AI system is intended to be used as a safety component of a product, or the AI system is itself a product, covered by the Union harmonisation legislation listed in Annex I};
\item
\textit{the product whose safety component pursuant to point (a) is the AI system, or the AI system itself as a product, is required to undergo a third-party conformity assessment, with a view to the placing on the market or the putting into service of that product pursuant to the Union harmonisation lgislation listed in Annex I}.

\end{enumerate}

Section 2, on the other hand, governs the \emph{requirements for
high-risk AI systems}, whose rules regulate such systems in greater
depth\emph{.}

The entire framework contained in the European Artificial Intelligence
Act can be characterized as ``\emph{human-centric}" because the European
legislature paid particular attention to the protection of people.

It is expressed precisely in Article 14, headed ``\emph{Human oversight}"
on high-risk artificial intelligence systems, as will be clarified in
the following section.

\section{Human Oversight}

As anticipated, Article 14(1) and (2) of the AI Act stipulate the
following:

\begin{enumerate}
\def\labelenumi{\arabic{enumi}.}
\item
  \begin{quote}
  \emph{High-risk AI systems shall be designed and developed in such a
  way, including with appropriate human-machine interface tools, that
  they can be effectively overseen by natural persons during the period
  in which they are in use.}
  \end{quote}
\item
  \begin{quote}
  \emph{Human oversight shall aim to prevent or minimise the risks to
  health, safety or fundamental rights that may emerge when a high-risk
  AI system is used in accordance with its intended purpose or under
  conditions of reasonably foreseeable misuse, in particular where such
  risks persist despite the application of other requirements set out in
  this Section.}
  \end{quote}
\end{enumerate}

The topic is central and particularly sensitive.

Specifically, the European legislature in Article 14(3) stipulates that
for high-risk AI systems, ``oversight measures" must be ensured through
the adoption of at least one of the two types declined in (a) and (b),
addressed to the provider or deployer, respectively.

But what does \emph{human oversight} consist of?

High-risk AI systems require governance expressed in the decision-making, as will become clear later.

Human oversight constitutes an approach aimed at ensuring human intervention (and here we refer to the \emph{human-centric} criterion mentioned above) in high-risk systems to
prevent the entire process from being governed by the artificial intelligence system.

Human \emph{oversight} is well known globally, so much so that doctrine has outlined different modes of approach, as will become clear later.

In the European context, already the document ``\textit{Ethics Guidelines for Trustworthy AI}" \cite{guidelines}, prepared by the Independent High-Level Expert Group on Artificial Intelligence set up by the European Commission in June 2018, presented the topic of ``Human Oversight" by proposing three governance mechanisms that allow for a \textbf{\textit{human-in-the-loop (HITL)}}, \textbf{\emph{human-on-the-loop (HOTL})} or \textbf{\emph{human-in-command} (\emph{HIC})} approach.

Moreover, in 2019, the document ``\textit{Communication from the Commission to the European Parliament, the Council, the European Economic and Social Committee and the Committee of the Regions Building Trust in Human-Centric Artificial Intelligence}" \cite{2019}, which outlines the European Artificial Intelligence Strategy, specifies that ``\textbf{Oversight can be carried out through governance mechanisms that ensure the adoption of an approach with human intervention (``human-in-the-loop"), with human supervision (``human-on-the-loop") or with human control (``human-in-command")}", thus reiterating the same methodological approach envisaged in the Expert Group paper as mentioned above.

Most recently, on March 20, 2024, a European Commission document, prepared together with the countries and stakeholders represented in the ERA Forum, entitled ``\textit{Living guidelines on the Responsible use of Generative AI in Research}" \cite{2024} was published, which contains simple and actionable directions to the European research community to promote the adoption of technology in a responsible manner. In this document, recalling the contribution entitled ``Ethical Guidelines for Reliable AI", it is emphasized that the four ethical principles for AI systems are:

\begin{enumerate}
\def\labelenumi{\arabic{enumi}.}
	\item \textit{respect for human autonomy};
	\item \textit{prevention of harm};
	\item \textit{fairness};
	\item \textit{explicability}.
\end{enumerate}

That document reiterates the same approach already set out in the others already mentioned, namely that ``\emph{human action and supervision includes human-in-the-loop (HITL), human-on-the-loop (HOTL), and human-in-command (HIC) approaches}".

As mentioned above, \emph{human oversight} is globally known, and the proposed approaches are closely related to "decision-making". In doctrine \cite{Singh}, the topic of decision-making is described in three different approaches related to varying degrees of artificial intelligence, namely, ``automated decision", ``decision augmented," and ``decision supported".

\emph{Human} oversight connected to decision-making has been developed with different approaches and used as models in machine learning.

According to a recent contribution \cite{Ivanov}, there are four approaches of \emph{human oversight} in decision \emph{making} (\emph{decision making}) related to the artificial intelligence field, namely:
\begin{enumerate}
\def\labelenumi{\alph{enumi})}
\item
  \emph{\textbf{Human-only approach}} - AI is not used in
  decision-making.
\item
  \emph{\textbf{Human-in-the-loop (HITL)}} - AI recommends a decision
  but it is up to humans to make it or not.
\item
  \emph{\textbf{Human-on-the-loop (HOTL)}} - AI makes and implements a
  decision, but humans can override it.
\item
  \emph{\textbf{Human-out-of-the-loop (HOOTL)}} - AI makes and
  implements decisions without human beings being able to change them.
\end{enumerate}

HHowever, other authors \cite{Singh} advocate a partially different classification of \textit{human oversight}, which is as follows:
\begin{enumerate}
\def\labelenumi{\alph{enumi})}
\item
 \textbf{\textit{Human-in-the-loop (HITL)}}- Humans make the decision,
 nd AI only provides support in making the decision.
\item
 \textbf{\textit{Human-in-the-loop-for-exceptions (HITLFE)}}- Most
 ecisions are automated, and humans only handle exceptions.
\item
 \textbf{\textit{Human-on-the-loop (HOTL)}}- The AI part makes
micro-decisions, and humans only assist the AI.
\item
 \textbf{\textit{Human-out-of-the-loop (HOOTL)}}- The AI part makes
 very decision; the human intervenes only to set new goals,
 onstraints, etc.
\end{enumerate}

Other authors\ \cite{Lena}, n the vein of the approach adopted by Europe, use a classification
 hat takes into account only the models usable in Machine Learning, \textit{\textbf{Human-in-the-loop (HITL)}},  \textit{\textbf{Human-on-the-loop (HOTL)}}, and \textit{\textbf{Human-In-Command (HIC)}}.

In the ``\textbf{Human-in-the-Loop (HITL)}" model \cite{nunes}, humans are directly involved in the development and algorithmic operations of artificial intelligence systems through supervisory activities. With the HITL model, the activities performed by humans mainly involve annotating so-called raw data to make labeled datasets needed for learning. In addition, humans provide feedback on the outcome of verifying results, pointing out any errors, and providing corrections and improvements. In essence, humans supervise the algorithms. HITL is, therefore, based on human interaction, intervention, and judgment aimed at controlling or modifying the outcome of a process using machine learning and generative AI in artificial intelligence systems.

More specifically, but without wishing to elaborate, the HITL approach involves humans performing controls in artificial intelligence systems' training, testing, deployment, and monitoring phases.

In essence, active control of the human being would reduce the risks of bias.
UUnlike the HITL model, systems using the ``\textbf{Human-on-the-Loop (HOTL)}" model \cite{Fischer} have greater autonomy, although they still involve human supervision. Human intervention is reduced to only cases of malfunction, serious errors, or other causes that profoundly affect the result (output). This model is usually used with extensively tested and reliable artificial intelligence systems.

Finally, the Machine Learning \emph{\textbf{Human-In-Command (HIC)}} model is one in which a human has ultimate authority and responsibility over an AI system.

Beyond the considerations made so far, a part of the doctrine \cite{benanti} has pointed out that human surveillance must integrate an ethical approach into the human-in-the-loop model. More specifically, said doctrine proposes a new governance that considers ethical aspects, calling it "algoretic" precisely as a fusion of the algorithmic and ethical components \footnote{The author affirms ``\emph{The insight into the need to create bodies or institutions that ensure the governance of technologies related to artificial intelligence seems very important. Only by making institutional places where these forms of ethical dialogue and regulation of technologies can take place can a real objective search for the good be addressed.}".}.

\section{Large Language Models (LLMs) - Introduction}

To introduce ``\emph{Large Language Models}" (LLMs), it is necessary to briefly describe what is called "foundation models" in the field of artificial intelligence.

``\emph{\textbf{Foundation}} models" (FM) are AI models designed to produce a broad and general variety of outputs such as text, images, audio, etc.

The phrase ``\emph{foundation models}" was coined \cite{Rishi1} by the Stanford Institute for Human-Centered Artificial Intelligence in the paper titled ``\emph{On the Opportunities and Risks of Foundation Models}" and described in these terms: ``\emph{any model that is trained on broad data (generally using self-supervision at scale) that can be adapted (e.g., fine-tuned) to a wide range of downstream tasks}".

According to a study by eight artificial intelligence researchers from the Center for Research on Foundation Models (CRFM) and the Institute on Human-Centered Artificial Intelligence (HAI) at Stanford University, as well as the MIT Media Lab and the Center for Information Technology Policy at Princeton University \cite{Rishi2}, \emph{foundation models} include ``generative AI" (GAI) and ``large language models" (LLM).

Beyond the origin of the expression "\emph{foundation models}" as just stated, there seems to be no unambiguity in the definition, which is often used interchangeably with GPAI (General-Purpose AI).

The European Artificial Intelligence Act regulates ``General-Purpose AI (GPAI) Models" in Chapter V (the `\emph{Classification of general-purpose AI models as general-purpose AI models with systemic risk}" is given in Article 51 and Annex XIII.

Article 3(63) defines ``\textbf{general purpose AI model}" in the following terms:

\begin{quote}
\emph{`\textbf{general-purpose AI  model}’ means an AI model, including where such an AI model is trained with a large amount of data using self-supervision at scale, that displays significant generality and is capable of competently performing a wide range of distinct tasks regardless of the way the model is placed on the market and that can be integrated into a variety of downstream systems or applications, except AI models that are used for research, development or prototyping activities before they are released on the market.}
\end{quote}

Systemic risk, on the other hand, is defined in Article 3(65) as follows:

\begin{quote}
\emph{`\textbf{systemic risk}’ means a risk that is specific to the high-impact capabilities of general-purpose AI models, having a significant impact on the Union market due to their reach, or due to actual or reasonably foreseeable negative effects on public health, safety, public security, fundamental rights, or the society as a whole, that can be propagated at scale across the value chain.}
\end{quote}

In essence, ``foundation models" can be described as the general category of AI systems developed to be adapted for a variety of different and specific purposes; they are the model that forms a base (hence ``foundation") on which other things can be developed. This qualification distinguishes these models from other AI systems trained and used in specific ways.

Another important distinction is between \textit{\textbf{Generative Artificial Intelligence (GAI)}} and \textbf{Large Language Models (LLM)}.

\textbf{GAI} is a variant of AI that can generate various human-usable content, such as \emph{natural language} (\emph{natural language}) text, images, audio, code, and domain-specific data \cite{Bridgelall} \textbf{Large Language Models (LLM)} \cite{IBM} \footnote{According to IBM, ``\emph{In a nutshell, LLMs are designed to understand and generate text like a human, in addition to other forms of content, based on the vast amount of data used to train them. They have the ability to infer from context, generate coherent and contextually relevant responses, translate to languages other than English, summarize text, answer questions (general conversation and FAQs) and even assist in creative writing or code generation tasks".}} are a category of \emph{foundation models. GAI} \cite{Bridgelall} is a type of artificial intelligence system trained on textual data that can generate natural language responses to input or requests.

\subsection{LLMs: GPT from OpenAI}

LLMs are rapidly evolving.

Examples of LLMs or applications of ``foundation models" are GPT-4 from OpenAI, BERT from Google, Claude 3 from Anthropic, DALL E from Open AI, and Llama2 from Meta.

Probably the best-known model among LLMs is GPT (short for \emph{Generative} \emph{Pre-trained Transformers} ) from OpenAI, which uses Deep Learning; it should not be confused with ChatGPT (short for ``\emph{Chatbot-Generated Text Predictor}") also from OpenAI, which instead is a chatbot built on GPT.

Therefore, the former (GPT) is a model of LLM, while ChatGPT is a chatbot.

The development of the GPT model began in 2018, and the current version is version 4 (GPT-4).

According to the paper titled ``\emph{GPT-4 Technical} Report" \cite{GPT}, ``\emph{GPT-4 is a Transformer-style model pre-trained to predict the next token in a document, using both publicly available data (such as internet data) and data licensed from third-party providers.}"

OpenAI\textquotesingle s GPT model has evolved from powerful \emph{text-based} tools (albeit with some limitations), such as GPT-3.5, to multimodal systems designed to perform numerous tasks, such as GPT-4.

The differences are substantially in the number of parameters (175 billion for GPT-3.5 and one trillion for GTP-4) and the number of tokens, which represent common character sequences\footnote{For details see \url{https://platform.openai.com/docs/introduction/tokens}} (16,385 tokens until September 2021 for gpt-3.5-turbo-0125 and 128,000 tokens until December 2023 for gpt-4-0125-preview)\footnote{Detailed information is available in the OpenAI documentation - \url{https://platform.openai.com/docs/models/overview}}.

The multimodal system (GPT-4) makes it possible to respond to requests that are not just text, producing content that can be multimedia (audio and video), graphics, images, etc.

\subsection{LLMs: limits, risks, and critical issues}

After this brief descriptive introduction, one cannot overlook the limitations, risks \cite{Tianyu}, and critical issues of LLMs\footnote{For an examination of the risks of LLMs, see \cite{Tianyu}.}, which, for the output part, may consist
of violation of \emph{copyright} legislation (\emph{copyright}), errors
in content rather than grammatical or syntactical errors, and phenomena
termed ``\textbf{hallucinations}".

On the other hand, LLMs were trained by having them acquire vast amounts of data in the form of texts. Given the disproportionate amount of data acquired, allegations arose of copyright infringement, data protection, and privacy regulations. It was suggested that LLM had acquired personal data from the Internet without the knowledge of the data subjects, and
this led the Italian Data Protection Authority (Garante) to issue a measure on March 30,
2023\footnote{\url{https://www.garanteprivacy.it/web/guest/home/docweb/-/docweb-display/docweb/9870832}}, ordering, as a matter of urgency, against OpenAI, the measure of provisional limitation, of the processing of personal data of data subjects established in Italian territory. Subsequently, the Italian Data Protection Authority (Garante), with a measure dated April 11, 2023\footnote{\url{https://www.garanteprivacy.it/web/guest/home/docweb/-/docweb-display/docweb/9874702\#english}}, ordered OpenAI to adopt the measures indicated in the abovementioned
resolution.

Regarding the user\textquotesingle s use of the outputs, one of the risks is not so much in the grammatical and syntactic part but precisely in the content of the response produced by the model.

It may be that LLMs provide an answer that is entirely wrong or out of context.

In particular, in the LLM context, the phenomenon of ``\textbf{hallucinations}" is well known, and it refers to hypotheses in which the model generates incorrect, nonsensical, or unreal. In
essence, LLM models generate text that results from not so much extrapolation from the data provided for training but is closely related to the prompt. 

Hallucination in LLMs arises from the process of abstraction whereby the model transforms the application for the data provided for training, and some information may be lost. Another aspect to consider in the phenomenon of ``hallucinations" is determined by the possible creation of a distorted statistical model caused by the ``noise" in the training data.

Someone highlights that ``the study of hallucinations in LMs is still in its infancy, even outside the legal field" \cite{dahl}.

\textbf{Biases} are one of the risks of artificial intelligence and can consist of biases or discrimination present in the output of an AI system. 

Biases can be different (social, racial, etc.) with obviously severe consequences. Sometimes, it depends on the data used for training
or the algorithmic choices made.

Another risk in LLMs is \textbf{adversarial prompting,} which is actual user attacks delivered through specific techniques designed to bypass the security barriers of the model. These cases fall under ``\emph{prompt engineering}", and the known types of attacks are prompt injection,
prompt leaking, jailbreaking, and DAN (short for ``Do Nothing Now").

It is no coincidence that OpenAI\textquotesingle s GPT-4 page reads, ``\emph{GPT-4 still has many known limitations that we are trying to address, such as social biases (}social biases\emph{), hallucinations, and} adversarial prompts".

Privacy and \emph{data protection} risks should also be added to the abovementioned risks.

\subsection{LLMs and the legal profession}

Although LLMs are relatively recent, there is already no shortage of some precedents related to applying outputs in the legal field with embarrassing consequences.

\textbf{8-22/6/2023} - The first known case is that of Roberto
Mata \cite{Mata}, who was struck by a metal service cart aboard a 2019 Avianca Airlines
flight and brought an action for damages. Mata\textquotesingle s lawyers cited at least six other cases as precedents. Still, the court deemed them nonexistent (a phenomenon known as ``hallucinations"), prompting a federal judge to consider sanctions. It later emerged that attorney Steven Schwartz-who represented Mata before proceeding to trial (he was not qualified to defend before the court)-admitted to using ChatGPT, claiming that he did not understand that it was not a search engine but a generative language processing tool. U.S. District Judge P. Kevin Castel of Manhattan ordered attorneys Steven Schwartz, Peter LoDuca, and their law firm Levidow, Levidow \& Oberman to pay a total fine of \$5,000 \cite{reuters}.

Other cases involving the use of ChatGPT are given below.

\textbf{22/2/2024} - The Cuddy Law Firm has asked the city of New York to pay its legal fees (over \$113,000 plus interest) quantified by ChatGPT in a successfully settled court case. However, U.S. District Judge Paul Engelmayer awarded just over \$53,000 and criticized the firm for using the artificial intelligence tool ``as support for its aggressive fee is absolutely and unusually unconvincing" \cite{Bloomberg}.

\textbf{20/3/2024} - A federal judge has refused to sanction Michael Cohen, Donald Trump\textquotesingle s former fixer, for mistakenly providing his lawyer (David Schwartz) with fake case citations generated by Google Bard (now Gemini), calling the incident ``embarrassing" \cite{Cohen}.

\textbf{26/2/2024} - A B.C. Supreme Court judge rebuked lawyer Chong Ke \cite{cbc} for including two AI ``hallucinations" in an application filed last December on behalf of Chen to obtain an order allowing his children to travel to China. Judge David Masuhara said he did not think the lawyer had intended to mislead the court but was nonetheless troubled, pointing out, ``\emph{As this case has unfortunately made clear, generative AI cannot yet replace the professional competence that the court system requires of lawyers}."

The four cases cited above demonstrate how dangerous and professionally disqualifying it is to use systems such as ChatGPT or other LLMs for work purposes without prior verification of the output.

Beyond severe professional censure, it is profound naiveté that denotes how poor awareness of artificial intelligence systems and knowledge of LLMs are; lawyers and professionals believe they can use (for free) such systems to deal with professional matters.

In the past, many ignored (or were fully aware of) that the unrestricted use of LLMs constituted system training, that is, the user trained the artificial intelligence system.

\subsection{LLMs and Human Oversight.}

LLMs are not free from risks, as summarized in the preceding paragraphs.

Unconditional use of output can generate seriously prejudicial situations, which has been documented in use cases by some lawyers.

However, hallucinations or biases can produce a result (output) that would still be incorrect or distorted concerning reality, and its use is irrespective of the fields, professional or otherwise, in which it is used.

It has been said that LLMs are not without risks related to personal data protection and privacy, which can have serious repercussions.

Therefore, although human oversight is a regulatory requirement under Article 14 of the European Artificial Intelligence Act (AI Act) for high-risk AI systems, it is not ruled out that it should also be used for LLMs.

LLMs could also be used within other AI systems, resulting in increased risk, mainly when employed in sensitive areas such as health care or in contexts where biometric data are used.

In such cases, human oversight, including through the adoption of specific policies, would be necessary beyond whether the AI system can be classified as high-risk.

\subsection{LLMs and Ethical Oversight}

Ethics is an indispensable component, especially in a context such as artificial intelligence.

European lawmakers have considered ethics so much that the Artificial Intelligence Act requires compliance with ethical principles, explicitly citing the ``\emph{Ethical Guidelines for Reliable AI developed by the independent AI HLEG.}" \cite{EComm}.

Moreover, Ethical principles are also mentioned in the recent document above entitled ``\emph{Living guidelines on the responsible use of generative AI in} research."\footnote{See footnote 8}.

As made clear in other contributions \cite{fabiano1} \cite{fabiano2}, ethics is a crucial element that helps to properly assess the nature of the impact, including legislative impact, of processes and enhance them through applying relevant principles.

More specifically, in the context of artificial intelligence (but this applies to any field, including data protection and privacy), AI systems may also affect more vulnerable or weak categories of people, which should be considered and not ignored. However, ethical evaluations should still be regarded as fundamental even in contexts with no vulnerable individuals or categories, precisely because some effects resulting from AI systems could also be considered potentially non-virtuous.

Therefore, without wishing to elaborate on this issue here, it is believed that ethical oversight is also necessary.

Ethical evaluations, however, involve checks and supervision that should be the responsibility of an ethics committee. In this regard, it is worth mentioning the work done by ForHumanity \cite{FH}, which has developed certification schemes focused on ensuring ethical oversight that mitigates bias for protected categories, intersectionality, and vulnerable populations.

\subsection{LLMs and the Data Protection Impact Assessment (DPIA)}

The impact that LLMs can also have on personal data leads one to believe
that it may be appropriate, if not necessary in the presence of the
conditions indicated by the GDPR, to carry out a \emph{Data Protection
Impact Assessment} (\emph{DPIA}), governed by Articles 35 and 36 of the
same GDPR. It is just worth noting that this institution enjoys an
autonomous place in Regulation 2016/69, given that it is placed in the
\emph{ad hoc} Section, headed - precisely - ``\emph{Data Protection
Impact Assessment and Prior Consultation.}"

Specifically, Article 35(1) and (2) of the GDPR states the following:

\begin{quote}

\emph{1. The Commission shall assign a single identification number to
each notified body, even where a body is notified under more than one
Union act.}

\emph{2. The Commission shall make publicly available the list of the
bodies notified under this Regulation, including their identification
numbers and the activities for which they have been notified. The
Commission shall ensure that the list is kept up to date.}

\end{quote}

Four key aspects emerge from the cited standard, namely:

\begin{enumerate}
\def\labelenumi{\alph{enumi})}
\item
  The use of \textbf{new technologies} in processing.
\item
  The possibility that the processing presents a \textbf{high risk} to
  the rights and freedoms of natural persons.
\item
  The owner must conduct an \textbf{impact assessment} before
  processing.
\item
  When conducting an impact assessment, the owner must \textbf{consult
  with the DPO}.
\end{enumerate}

The first two conditions stated in the cited standard (paragraph 1)
elicit comments aimed at seeking clarification. More specifically, the
expression ``\textbf{new technologies}" appears ambiguous, as it would be
contrasted with `old technologies." It is unclear what is to be
understood by new or old technologies, assuming it is possible to
identify technologies that can be qualified as recent. The GDPR does not
clarify what the proper meaning to be given to the expression ``new
technologies," not even Recital (89), which - among other things -
states, ``Such types of processing include, in particular, those
involving the use of new technologies or those which are of a new kind
... is" The Recital mentioned above, therefore, does not provide
valuable hermeneutical support to clarify what the meaning of ``new
technologies" is, nor does the following expression ``or those {[}Ed:
treatments{]} that are of a new kind" prove helpful. At the outcome of a
hermeneutic investigation, it should be considered that technologies
that are newly discovered or developed, such that they have, even if
only potentially, a significant impact in the different areas of the
life of every human being and, therefore, qualify as emerging or
innovative, are to be qualified as `new." The element that would
characterize "new technologies" in terms of novelty would, therefore, be
the special significance of both a technological nature and of the
current or future impact on the life of every human being.

In this regard, Recital (91) of the GDPR, which states:

\begin{quote}

\emph{`` \ldots{} which are likely to result in a high risk, \textbf{for
example, on account of their sensitivity, where in accordance with the
achieved state of technological knowledge a new technology} is used on a
large scale as well as to other processing operations which result in a
high risk to the rights and freedoms of data subjects, in particular
where those operations render it more difficult for data subjects to
exercise their rights''}

\end{quote}

It is clear, therefore, that the criterion used by the European
legislator is that of the ``degree of technological knowledge attained,"
according to which novelty is to be parameterized to the current
condition; that which is not known technologically constitutes the
element by which the term "new" can be attributed.

In light of this, it is not possible to consider, for example, the
Internet per se as a ``new technology" since today it does not take on
the character of novelty, both about the technological profile and
concerning the impact on the lives of human beings already realized in
the past. It has been said, however, that the Internet itself today
cannot be considered a ``new technology," but the same cannot be said for
the technologies used for its evolution. It is well known that the web
has evolved through technologies (Web 3, blockchain, etc.) that have
changed and will continue to change communication. Thus, the
discriminator that could qualify a technology as ``new" is its
topicality, its characteristics of novelty, experimentation, innovation,
and its potential further development that will impact a change in the
life of every human being.

Another aspect that deserves to be commented on and necessarily deepened is the concept of ``\textbf{high risk}", in light of the same classification criterion (high-risk) used by the AI Act for artificial intelligence systems. 

The expression ``high risk" is used several times in the GDPR, and the Recital above (91) exemplifies high risk with the hypothesis that ``a new technology is used on a large scale", which may impact the rights and freedoms of the data subjects.

Beyond hermeneutic investigations, the processes and tools helpful in identifying the risk, the relative level at which it can be qualified (low, medium, or high), and its management are also relevant in practice. The GDPR does not indicate which criteria to adopt to identify, qualify, and manage risk; therefore, one must use the aforementioned international standards for risk management and risk assessment (ISO/IEC 31000 and ISO/IEC 31010).

The AI Act does not classify LLM as a high-risk artificial intelligence system. However, from the above, it is clear that it can also be a source of high risk depending on its implementation in an organization's processes, the conditions under which it is used, and the use of its outputs.

We have described some consequences of careless use of the results obtained by LLM users (ChatGPT and Bard) in a sensitive professional field.

LLM, however, was not created solely for anyone's free use but for obvious business reasons. That being said, there is no doubt that LLM can be implemented within artificial intelligence or other algorithmic systems by using developers' APIs (Application Programming Interfaces) that allow different applications to communicate. Consider, for instance, the provisions of the European Digital Markets Act precisely in this area to encourage the use of applications of so-called gatekeepers also by others by allowing them to communicate their algorithmic solutions.

Solutions of this kind could raise the level of risk to such an extent that LLM in their applications and implementations would have to be considered high-risk. 

We cannot dwell on this here, but another aspect that could affect LLM's output is metadata, both in the training phase and as pure output, i.e., as results extracted, for instance, from documents.

In light of the foregoing, in the field of personal data protection, an impact assessment must always be required, both in the presence of high-risk AI systems and of algorithmic systems or solutions implementing LLM. From the assessment emerges an increase in the level of risk that cannot be mitigated with specific measures (e.g., because there are no effective solutions in the state of the art).

Indeed, Article 36 of the GDPR, entitled Prior Consultation, states the following in paragraph 1:

\begin{quote}
	
\emph{``1. The controller shall consult the supervisory authority prior
to processing where a data protection impact assessment under Article 35
indicates that the processing would result in a high risk \textbf{in the
absence of measures taken by the controller to mitigate the
risk}}".

\end{quote}

Where the data controller uses high-risk AI systems, the abovementioned GDPR will have to be coordinated with those of the AI Act related to the risk management system. 
One should carry out the same coordination whenever faced with complex LLM, possibly implemented in other AI systems with high risks that are difficult to mitigate.
These complex activities entail full regulatory knowledge and mastery of the technologies used; only the combination of these two fundamental elements makes it possible to initiate the appropriate assessment and compliance processes.
Therefore, in the cases indicated, in the absence of measures taken by the data controller to mitigate the risk, prior consultation is always necessary. 
In these contexts, the DPO plays a key role.

\section{Artificial intelligence Liability}

Some critical issues related to the possible prejudicial consequences of
artificial intelligence systems have been described. Damage may result
from the injury; thus, the topic debated whether AI liability can be
attributed to it, i.e., who is the party liable to compensate for the
damage.

The question depends, of course, on the legal system and, thus, for
Europe, on a regulatory source governing liability arising from
artificial intelligence and for each member state on national
legislation.

In Europe, the European Commission published the following two proposals
on September 28, 2022:

\begin{enumerate}
\def\labelenumi{\alph{enumi})}
\item
\textbf{Proposal for a Directive of the European Parliament and of the Council on liability for defective products (Product Liability Directive - PLD Proposal)}\footnote{The last update is that the European Parliament adopted the text at first reading on 12 March 2024} \cite{PLD};
\item
\textbf{Proposal for a Directive of the European Parliament and of the Council on adapting the rules on non-contractual civil liability to artificial intelligence (Artificial Intelligence Liability Directive) (AILD Proposal)} \cite{AILD}.
  
\end{enumerate}

The European Data Protection Supervisor (EDPS) has expressed its views
in EDPS Opinion 42/2023 on the Proposal for two Directives on AI
liability rules \cite{EDPS}, and from the conclusions of which the main following aspects emerge:

(3) \textbf{Extend the procedural safeguards} provided in Articles 3 and
4 of the AILD proposal, i.e., disclosure of evidence and presumption of
causation, to all cases of harm caused by an AI system, regardless of
its classification as high-risk or non-high-risk;

(4) \textbf{Ensure that information disclosed under} Article 3 of the
AILD proposal is accompanied by explanations to be provided in an
intelligible and generally understandable form;

(6) \textbf{Consider} further \textbf{measures} to further ease the
burden of proof for \textbf{victims of AI systems} as a means of
ensuring the effectiveness of EU and national liability rules.

The topic is discussed because, from a legal point of view, it would not
be permissible to recognize the responsibility of a machine.

Indeed, liability is attributable to a human being, but legal systems do
not regulate the liability of machines or artificial intelligence
systems. 

Artificial intelligence systems operate and behave as they have
always been designed and programmed by human beings. 

We do not rule out
the possibility of artificial intelligence systems or machines being
designed and programmed by other machines.

Therefore, It is clear that the issue of liability for AI systems is
currently open and subject to debate. Moreover, the EDPS opinion
emphasizes some issues in the proposed directives: burden of proof,
presumption of causation, damage caused by an AI system, precise
information regarding AI systems, and effective enforcement of European
and national rules on the liability of AI systems.

If the issue of liability of AI systems can be addressed without
particular constraints at the European level due to its characteristic
supranational, some critical issues might be critical in national areas
where the nature of the legal system regulates liability. 

In essence, should the abovementioned directive proposals be approved, an obligation
would arise for individual member states to transpose them. At the transposition stage, the national legislator will have to consider the legal nature of the institution of liability, its location, and the \emph{rationale of} the regulation. 

The introduction at the national level of the liability of a system (of IA) or a machine could conflict not only with the existing discipline but precisely with the legal
qualification of the liability itself, which could be incompatible - by its nature and underlying rationale - with the legal system. All this would entail a profound change in civil and criminal law\textquotesingle s entire institution of liability.

\section{Conclusions}

Large Language Models (LLM), artificial intelligence systems that can be
characterized as "foundation models" or \emph{Generative Artificial
Intelligence} (GAI), are rapidly evolving.

LLMs are not without risks, as discussed, albeit briefly.

Because of their architecture, for which Machine Learning, Deep
Learning, and \emph{Pre-trained Transformers} models are used on vast
amounts of data that are used for training, \emph{human oversight}
(\emph{human oversight}) proves necessary, whether the output is used
directly by users or if it is usable by other AI system.

By their structure, LLMs would not allow the use of Machine Learning
models such as \emph{Human-in-the-loop}, \emph{Human-on-the-loop,} or
\emph{Human-in-command}, partly because they are not considered
high-risk artificial intelligence systems.

As discussed above, the ethical component is crucial, and \emph{ethical
oversight} must also be provided.

LLMs are not exempt from data protection impacts; therefore, impact
assessments (DPIAs) should be conducted to assess any high risks to the
rights and freedoms of individuals.

\cleardoublepage

\end{document}